# Tip-enhanced Raman spectroscopic detection of aptamers

Siyu He[1,2], Hongyuan Li[1,2], Zhe He[1], and Dmitri V. Voronine[1,3]

[1] Texas A&M University, College Station, TX 77843, USA
[2] Xi'an Jiaotong University, Xi'an 710049, China
[3] Baylor University, Waco, TX 76798, USA

**Abstract**

Single molecule detection, sequencing and conformational mapping of aptamers are important for improving medical and biosensing technologies and for better understanding of biological processes at the molecular level. We obtain vibrational signals of single aptamers immobilized on gold substrates using tip-enhanced Raman spectroscopy (TERS). We compare topographic and optical signals and investigate the fluctuations of the position-dependent TERS spectra. TERS mapping provides information about the chemical composition and conformation of aptamers, and paves the way to future single-molecule label-free sequencing.

## 1. Introduction

Surface- and tip-enhanced Raman scattering can be used to reveal the molecular bonds as well as the functional components in biomaterials, which was applied to few and single molecule sensing[1,2,3,4]. Surface-enhanced Raman scattering (SERS) utilizes plasmonic resonances of metallic nanostructures to enhance Raman signals of biomolecules[5]. Weak Raman signals of complex biomolecules could be challenging for measurements at the single molecule level[6,7]. SERS can be used boost the Raman signals at the plasmonic resonance condition.

Tip-enhanced Raman spectroscopy (TERS) is scanning probe microscope combined with the plasmonic enhancement of SERS and provides nanoscale label-free, direct chemical imaging method due to high lateral resolution and possibility of simultaneous imaging and sensing at the single-molecule level[4,8]. Recently, several DNA sensing methods have emerged in various fields of research as subjects of intense studies[9,10]. Conventional DNA sensing utilized dye labelling and enzyme processing, which modify target molecules. On the other hand, enhanced Raman sensing provides minimally invasive label-free bio-sensing. Also, the tip and the substrate can have an additional gap-mode enhancement which provides better spatial resolution and stronger Raman signals[11].

Here, we report first TERS mapping of aptamers immobilized on gold substrates and the corresponding spectral variations and spatial resolution of the single molecule TERS signals.

## 2. Experimental methods

Thiol-functionalized aptamers for Listeria monocytogenes consisted of 47−unit DNA oligomers and are the targets for the internalin A (InlA) protein[12]. The sequence and schematic conformational structure of such aptamer is shown in Figs. 1a and 1b, respectively. These aptamers have both single and double stranded DNA parts in one molecule. Only the 3’ end was modified with a terminal thiol group which works as a constraint for the aptamer motion on the gold surface. This functionalization scheme may be used for the purification of DNA by adsorption to a metallic substrate[13]. Here we used a 1 cm × 1 cm atomically flat gold substrate for its suitability for single molecule topographic imaging and optical field gap mode enhancement.

The thiol functionalization of DNA was performed following the previously reported protocols[14]. The gold substrate was cleaned with piranha solution with the ratio of 3:1 concentrated sulfuric acid to hydrogen peroxide for one minute to remove all the organic residues from the surface of gold. Then the gold substrate was washed using deionized water for one minute to ensure the residues and piranha solution were removed. After that, the gold substrate was air dried. The stock solution of 13 $\mu$m Listeria monocytogenes DNA was diluted to 100 nM in water. Then 65 $\mu$L of the solution was used to functionalize onto the gold substrate by drop coating and air drying for 12 hours inside a biosafety cabinet. Following that, the gold substrate was washed using deionized water 3 times and air dried for 30 min.

TERS imaging experiments were performed using Omega-Scope (AIST-NT) instrument combined with the LabRam Raman microscope (Horiba), using a gold-coated nano-tip in the contact AFM mode. AFM imaging was performed using the tapping mode with 10 nm amplitude. The laser excitation wavelength was 660 nm. Gold-coated TERS tips were purchased from AIST-NT and had a radius of ~ 20 nm. The acquisition time of each spot in TERS mapping was 5 s.

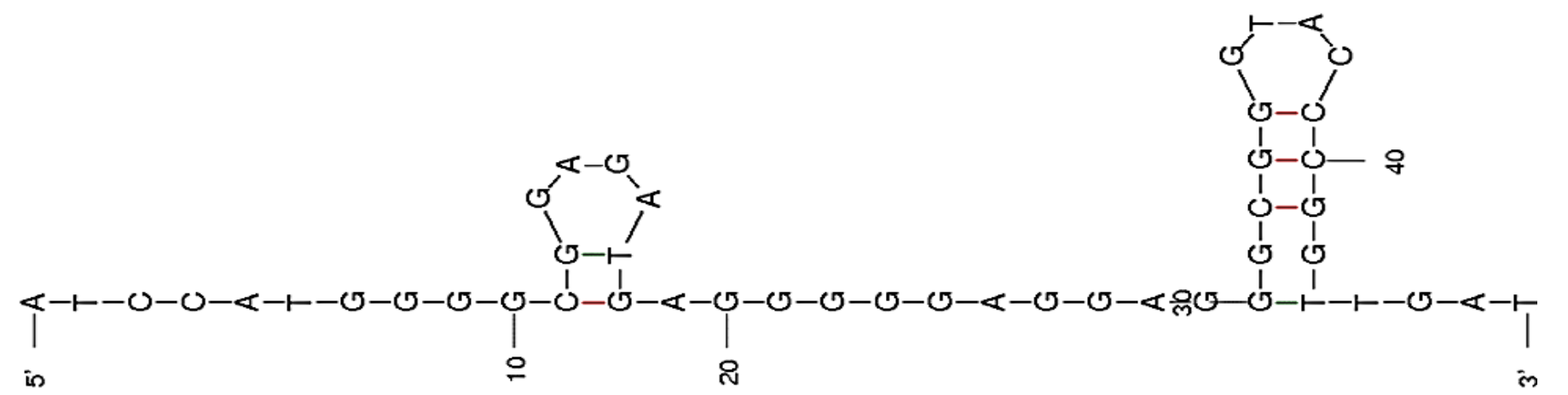

(a)

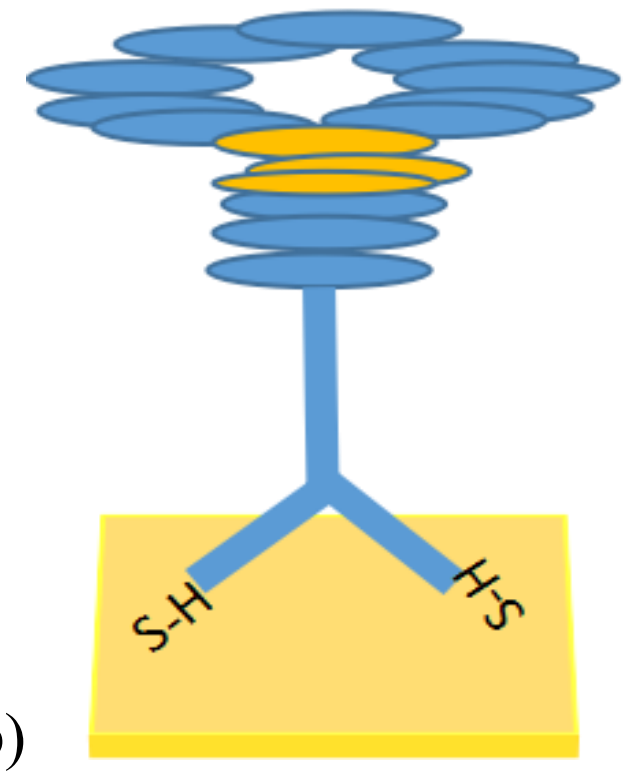

(b)

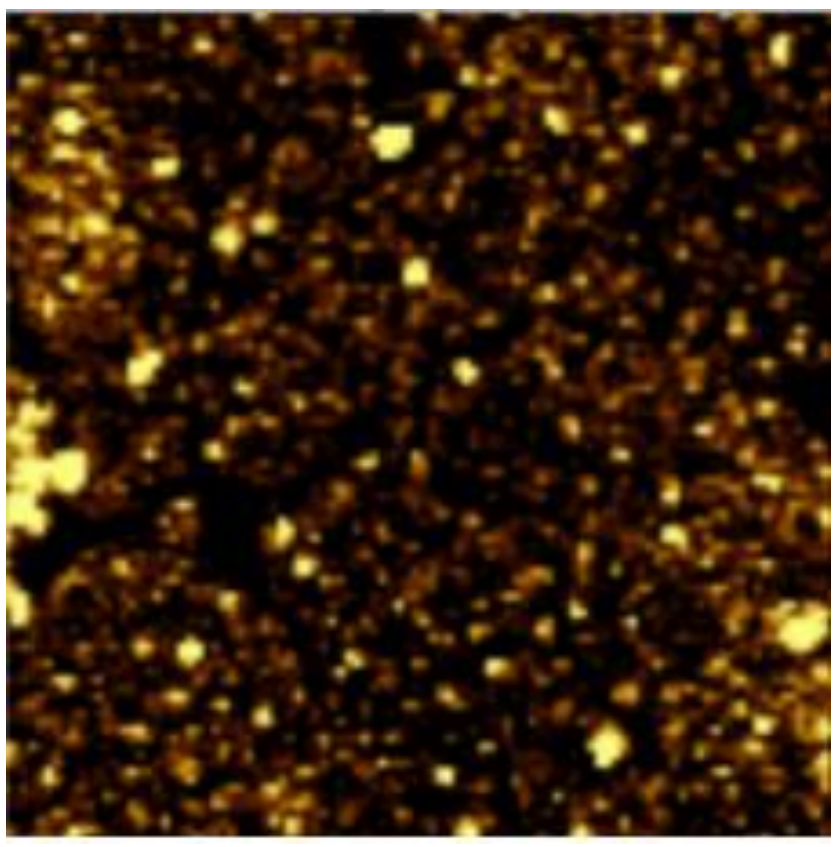

(c)

Figure 1. Sequence (a) and schematic structure (b) of aptamer for L. monocytogenes functionalized on the gold substrate. (c) AFM image of a large area on the gold substrate containing many aptamers.

## 3. Results and discussion

The largest dimension size of the naturally folded aptamer on the gold surface was about 20 nm with each nucleobase size of ~ 0.676 nm[15], whereas was measured to be about 40 nm in AFM image in Fig. 1c which was limited by the size of the gold tip. Fig. 1c shows an array of many aptamer molecules on a micrometer size area of the gold substrate. The height of the aptamers varies due to the differences in conformational configurations of individual molecules, their dynamics and varying substrate environment. In our AFM experiments most spherical spots showed in-plane topographical dimensions of ~ 40 nm and ~ 2 nm in height. Therefore, in the subsequent TERS analysis we selected the spots with 40 nm width and 2 nm height.

Fig. 2a is a micrometer size TERS map with a 10 nm x 10 nm pixel size which corresponds to the area characterized by AFM shown in Fig. 1c. The TERS map pixel size is of the same order of magnitude as the size of the folded aptamer. Therefore, it was not possible to obtained a detailed conformation and sequence from the TERS map with such spatial resolution. However, we obtained Raman signals from individual aptamer molecules which showed correlation with the topography. An enlarged view of the selected region of interest 3 (ROI3) in Fig. 2b shows several isolated hot spots with strong Raman signals. The corresponding spectra are shown in Fig. 2c and labeled according to literature-based band assignment. Fig. 2d shows a comparison of the TERS signals from two closely spaced locations which confirm reproducibility of the high resolution Raman maps.

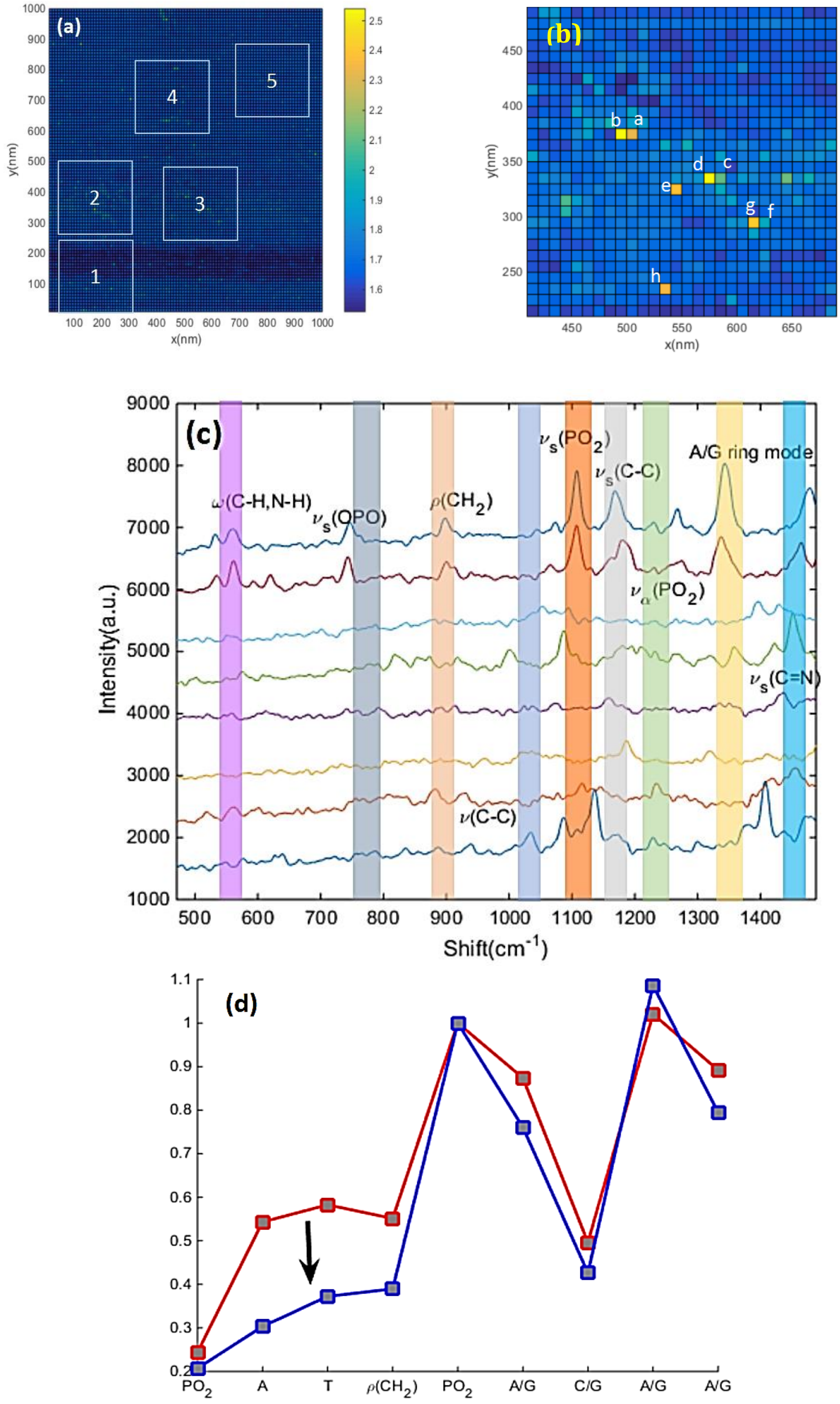

Figure 2. (a) TERS map which corresponds to the AFM map in Fig. 1c with five highlighted regions of interest (ROIs). (b) TERS of ROI3 with several hot spots marked by letters a – h with the corresponding Raman spectra arranged in the descending order from bottom to top, respectively (c). (d) Comparison of the reproducibility and band assignment of two closely spaced locations with strong Raman signals.